\documentstyle[aps,prd,tighten,epsf]{revtex}
\begin{document}

\title{Evolution of a Self-interacting
 Scalar Field in
the Spacetime of a Higher Dimensional Black Hole}

\author{ Rafa{\l} Moderski}
\address{
Nicolaus Copernicus Astronomical Center \protect \\
Polish Academy of Sciences \protect \\
00-716 Warsaw, Bartycka 18, Poland \protect \\
moderski@camk.edu.pl }

\author{Marek Rogatko}

\address{Institute of Physics \protect \\
Maria Curie-Sklodowska University \protect \\
20-031 Lublin, pl.Marii Curie-Sklodowskiej 1, Poland \protect \\
rogat@tytan.umcs.lublin.pl \protect \\
 rogat@kft.umcs.lublin.pl}

\date{\today}
\maketitle

\begin{abstract}
In the spacetime of n-dimensional static charged black hole
we examine the mechanism by which the self-interacting scalar hair decay.
It is turned out that the intermediate asymptotic behaviour of
the self-interacting scalar field is determined by an oscillatory inverse power law.
We confirm our results by numerical calculations.
\end{abstract}

\newcommand{\be}{\begin{equation}}
\newcommand{\ee}{\end{equation}}
\newcommand{\ben}{\begin{eqnarray}}
\newcommand{\een}{\end{eqnarray}}

\newcommand{\la}{{\lambda}}
\newcommand{\Om}{{\Omega}}
\newcommand{\ta}{{\tilde a}}
\newcommand{\bg}{{\bar g}}
\newcommand{\bh}{{\bar h}}
\newcommand{\si}{{\sigma}}
\newcommand{\th}{{\theta}}
\newcommand{\C}{{\cal C}}
\newcommand{\D}{{\cal D}}
\newcommand{\cA}{{\cal A}}
\newcommand{\cT}{{\cal T}}
\newcommand{\cO}{{\cal O}}
\newcommand{\eeo}{\cO ({1 \over E})}
\newcommand{\G}{{\cal G}}
\newcommand{\cL}{{\cal L}}
\newcommand{\T}{{\cal T}}
\newcommand{\M}{{\cal M}}

\newcommand{\p}{\partial}
\newcommand{\na}{\nabla}
\newcommand{\ssum}{\sum\limits_{i = 1}^3}
\newcommand{\dssum}{\sum\limits_{i = 1}^2}
\newcommand{\tal}{{\tilde \alpha}}

\newcommand{\tp}{{\tilde \phi}}
\newcommand{\tPhi}{\tilde \Phi}
\newcommand{\tpsi}{\tilde \psi}
\newcommand{\tim}{{\tilde \mu}}
\newcommand{\tr}{{\tilde \rho}}
\newcommand{\tir}{{\tilde r}}
\newcommand{\rp}{r_{+}}
\newcommand{\hr}{{\hat r}}
\newcommand{\rv}{{r_{v}}}
\newcommand{\dr}{{d \over d \hr}}
\newcommand{\dR}{{d \over d R}}

\newcommand{\hhf}{{\hat \phi}}
\newcommand{\hhM}{{\hat M}}
\newcommand{\hhQ}{{\hat Q}}
\newcommand{\hht}{{\hat t}}
\newcommand{\hhr}{{\hat r}}
\newcommand{\hhS}{{\hat \Sigma}}
\newcommand{\hhD}{{\hat \Delta}}
\newcommand{\hhm}{{\hat \mu}}
\newcommand{\hro}{{\hat \rho}}
\newcommand{\hhz}{{\hat z}}

\newcommand{\tD}{{\tilde D}}
\newcommand{\tB}{{\tilde B}}
\newcommand{\tV}{{\tilde V}}
\newcommand{\hT}{\hat T}
\newcommand{\tF}{\tilde F}
\newcommand{\tT}{\tilde T}
\newcommand{\hC}{\hat C}
\newcommand{\ep}{\epsilon}
\newcommand{\bep}{\bar \epsilon}
\newcommand{\ppp}{\varphi}
\newcommand{\Ga}{\Gamma}
\newcommand{\ga}{\gamma}
\newcommand{\hth}{\hat \theta}
\bigskip

\baselineskip=18pt
\par
\section{Introduction}
The problem of the late-time behaviour of various fields in the spacetime of a collapsing
body is well established (both analytically and numerically). It plays an important role
in black hole's physics. Black hole radiates away everything that it can. 
This phenomenon happens regardless of details of the collapse or the structure and properties
of the collapsing body. The resultant black hole can be described only by few parameters
such as mass, charge and angular momentum and due to the Wheeler's metamorphic dictum
{\it black holes have no hair} (see for the vast amount of references concerning this problem \cite{uniq}).
Therefore
it is interesting to investigate how these hair loss proceed dynamically. 
\par
The neutral external perturbations were first studied in Ref.\cite{pri72}. It was found that
the late-time behavior is dominated by the factor $t^{-(2l + 3)}$, for each 
multipole moment $l$. On the other hand, 
the decay-rate 
along null infinity and along the
future event horizon was governed
by the power laws $u^{-(l + 2)}$ and $v^{-(l + 3)}$, where $u$ and $v$ were
the outgoing Eddington-Finkelstein (ED) and ingoing ED coordinates \cite{gun94}.
The scalar perturbations on Reissner-Nordtr\"om (RN)
background for the case when  $\mid Q \mid < M$ has the following dependence on time $t^{-(2l + 2)}$,
while for $\mid Q \mid = M$ the late-time behavior at fixed $r$ is governed by
$t^{-(l + 2)}$ \cite{bic72}.
\par
It turns out that a charged hair
decayed slower than a neutral one \cite{pir1}-\cite{pir3}, while
the late-time tails in gravitational collapse of a self-interacting (SI)
fields in the background of Schwarzschild solution was reported by Burko \cite{bur97}
and in
RN solution at intermediate late-time was considered
in Ref.\cite{hod98}. At intermediate late-time for small mass $m$ the decay was dominated
by the oscillatory inverse power tails $t^{-(l +3/2)} \sin (m t)$. This analytic prediction
was verified at intermediate times, where $mM \le mt \le 1/(mM)^2$. In Ref.\cite{ja}
the nearly extreme RN spacetime was considered and it was found analytically that the
inverse power law behavior of the dominant asymptotic tail is of the form
$t^{-5/6} \sin (m t)$, independent of $l$. The asymptotic tail behaviour 
of SI scalar field
was also studied in Schwarzschild spacetime \cite{ja1}. The oscillatory
tail of scalar field has the decay rate of $t^{-5/6}$ at 
asymptotically late time.
The power-law tails in the evolution of a charged massless scalar field around a fixed
background of dilaton black hole was studied in Ref.\cite{mod01a}, while the case of a self-interacting
scalar field was elaborated in \cite{mod01b}.
\par
Nowadays it seems that that it is impossible to construct a consistent theory unifying 
gravity with other forces in Nature in four dimensions. The {\it no-hair} theorem
for $n$-dimensional static black holes is quite well established \cite{unn}. So it will be not amiss
to ask about the mechanism of decaying black hole hair in higher dimensional static
black hole case. The evolution of massless scalar field in the $n$-dimensional Schwarzshild
spacetime was determined in
Ref.\cite{car03}. It was found that for odd dimensional spacetime the field  decay had a
power falloff like $t^{-(2l + n - 2)}$, where $n$ is the dimension of the spacetime.
This tail was independent of the presence of the black hole.
For even dimensions the late-time behaviour is also in the power law form but in this case it is due to
the presence of black hole $t^{-(2l + 3n - 8)}$.
Gravitational perturbations of maximally symmetric black hole spacetime in higher dimensions were 
studied by Kodama {\it et al.} \cite{kod03}, while gravitational quasi-normal radiation of 
higher dimensional black holes was
elaborated in \cite{kon03}.
\par
In our work we shall consider and discuss the
SI scalar field behaviour in the spacetime of $n$-dimensional static charged black hole.
In Sec.II we gave the analytic arguments concerning
the intermediate behavior
of SI scalar field in the background of the considered black hole.
Then, in Sec.III we treated the problem numerically and check our analytical considerations. 
We conclude our investigations in Sec.IV.

\section{Massive scalar fields in n-dimensional spherically symmetric spacetime}
\label{sec1}
In our paper we shall investigate the evolution of SI (massive)
scalar 
field $\tpsi$ in a fixed spacetime
of a static electrically charged
$n$-dimensional black hole. The wave equation for the field is given in the form as follows:
\be
{}{}^{(n)}\na_{\mu} {}{}^{(n)}\na^{\mu} \tpsi 
- m^2 \tpsi^2 = 0.
\ee
where $m$ is assumed to be real.
\par
The metric of the external gravitational field will be given by
the static, spherically symmetric solution of equations of motion
derived from the action written as
\be
S = \int d^n x \sqrt{-{}{}^{(n)}g} \left [
{}{}^{(n)} R - F_{(n-2)}^2 \right ],
\ee
where we denoted the generalized $(n - 2)$-gauge form $F_{\mu_{1} \dots \mu_{n-2}}$
by the expression $F_{(n-2)} = dA_{(n-3)}$. Further, we assume that we have to do
with the only one nontrivial {\it electric} component of the $(n - 2)$-gauge form
as $A_{0 1 \dots n-2} = \phi (x)$. The metric of the spherically symmetric charged static 
$n$-dimensional black hole implies
\be
ds^2 = - \bigg (
1 - {2 M \over r^{n-3}} + {Q^2 \over r^{2 n - 6}} \bigg) dt^2 +
{dr^2 \over \bigg (
1 - {2 M \over r^{n-3}} + {Q^2 \over r^{2 n - 6}} \bigg)} + r^2 d \Omega_{n-2}^2,
\label{metric}
\ee
where $d \Omega_{n-2}^2$ is the line element of the unit $S^{n-2}$ sphere.
Next, we define the tortoise coordinates $y$ as
\be
dy = {dr \over \bigg (
1 - {2 M \over r^{n-3}} + {Q^2 \over r^{2 n - 6}} \bigg)}.
\ee
Thus, the metric (\ref{metric}) can be rewritten in the form as
\be
ds^2 = \bigg (
1 - {2 M \over r^{n-3}} + {Q^2 \over r^{2 n - 6}} \bigg) \bigg[
- dt^2 + dy^2 \bigg] +  r^2 d \Omega_{n-2}^2.
\ee
In
the spherical background each of the multipole of perturbation
field evolves separetly. Because of the fact that the scalar field is of the form
\be
\tpsi = \sum_{l,m} {1 \over r^{{n-2 \over 2}}} \psi_{m}^{l}(t, r) 
Y_{l}^{m}(\theta, \phi),
\ee
where $Y_{l}^{m}$ is a scalar spherical harmonics on the unit $(n-2)$-sphere and
$m$ denotes a set of $(n-3)$ integers $(m_{1} \dots m_{n-3})$ satisfying
$l \ge m_{n-3} \ge \dots \ge m_{2} \ge \mid m_{1} \mid$. In the process of this
one has the 
following equations of motion for each multipole moment
\be
\psi_{,tt} - \psi_{,yy} + V \psi = 0.
\label{mo}
\ee
The potential $V$ implies
\be
V = f^2(r) \bigg[ \bigg( {n - 2 \over 2} \bigg)
{1 \over r^{{n-2 \over 2}}} {d \over dr} 
\bigg( r^{{n\over 2} - 2}~ f^2(r) \bigg) + {l (l + n - 3) \over r^2} + m^2
\bigg],
\ee
where $f^2(r) = 1 - {2 M \over r^{n-3}} + {Q^2 \over r^{2 n - 6}} $.

In order to analyze the time evolution of SI scalar field in the background of
the considered black hole we shall use the spectral decomposition method
\cite{lea}.
The time evolution of SI scalar field may be written in the following form:
\be
\psi(y, t) = \int dy' \bigg[ G(y, y';t) \psi_{t}(y', 0) +
G_{t}(y, y';t) \psi(y', 0) \bigg],
\ee
for $t > 0$, where   the Green's function  $ G(y, y';t)$ is given by
\be
\bigg[ {\p^2 \over \p t^2} - {\p^2 \over \p y^2 } + V \bigg]
G(y, y';t)
= \delta(t) \delta(y - y').
\label{green}
\ee
Our main task will be to find the black hole Green function so
in the first step we reduce equation
(\ref{green}) to an ordinary differential equation.
To do it one can use the Fourier transform \cite{and}
$\tilde  
G(y, y';\omega) = \int_{0^{-}}^{\infty} dt  G(y, y';t) e^{i \omega t}$.
This Fourier's transform is well defined for $Im~ \omega \ge 0$, while the 
corresponding inverse transform yields
\be
G(y, y';t) = {1 \over 2 \pi} \int_{- \infty + i \ep}^{\infty + i \ep}
d \omega~
\tilde G(y, y';\omega) e^{- i \omega t},
\ee
for some positive number $\ep$.
The Fourier's component of the Green's function $\tilde  G(y, y';\omega)$
can be written in terms of two linearly independent solutions for
homogeneous equation as
\be
\bigg(
{d^2 \over dy^2} + \omega^2 - V \bigg) \psi_{i} = 0, \qquad i = 1, 2,
\label{wav}
\ee
The boundary conditions for $\psi_{i}$ are described by purely ingoing waves
crossing the outer horizon $H_{+}$ of the 
$n$-dimensional static charged black hole
$\psi_{1} \simeq e^{- i \omega y}$ as $y \rightarrow  - \infty$ while
$\psi_{2}$ should be damped expotentially at $i_{+}$, namely
$\psi_{2} \simeq e^{- \sqrt{\omega^2 - m^2}y}$ at $y \rightarrow \infty$.
\par
In order to find $\psi_{i}$ we consider the wave Eq.(\ref{wav}) of SI 
scalar field  and introduce an auxiliary variable $\xi$ in such a way that 
$\xi = f(r) \psi$.
So in terms of $\xi$ relation (\ref{wav}) can be written as follows:
\be
{d^2 \xi \over d r^2} - {f(r)_{,rr} \over f(r)} \xi + 
+ {\omega^2 \xi \over f^4(r)} - {\xi \over f^2(r)}
\bigg[ \bigg( {n - 2 \over 2} \bigg)
{1 \over r^{{n-2 \over 2}}} {d \over dr} 
\bigg( r^{{n\over 2} - 2}~ f^2(r) \bigg) + {l (l + n - 3) \over r^2} + m^2
\bigg].
\label{wav1}
\ee
Further on, we expand Eq.(\ref{wav1}) in power series of ${M \over r^{n-3}}$ and
${Q \over r^{n-3}}$, neglecting terms of order 
$ {\cal O} \bigg( ({\alpha \over r^{ n-3}})^2 \bigg)$. Then, we arrive at the following expression:
\be
{d^2 \xi \over d r^2} + \bigg[ \omega^2 - m^2 + 
{4 M \omega^2 - 2 M m^2 \over r^{n-3}}
- {\nu (\nu + 1) \over r^2} \bigg] \xi = 0,
\ee
where $\nu = l - 2 + {n \over 2}$.\\ 
If one further assumes that the observer and the initial data are in the 
region where $r^{n-3}\ll{M \over (Mm)^2}$ and one shall be interested in the 
intermediate asymptotic behavior of SI scalar field $r^{n-3}\ll t \ll
{M \over (Mm)^2}$, then we get
\be
{d^2 \xi \over d r^2} + \bigg[ \omega^2 - m^2
- {\nu (\nu + 1) \over r^2} \bigg] \xi = 0.
\label{wave}
\ee
As in four-dimensional case the scalar field perturbations on $n$-dimensional charged static black hole 
background does not depend on the spacetime parameters such as $M$ and $Q$. The perturbations in question 
depend on the scalar field parameter (mass of the field).
\par
The same procedure as described in Ref.\cite{hod98} leads us to the solution of the relation (\ref{wave})
(we refer the readers to this work). Thus, in our case the intermediate asymptotic behaviour
of the SI field at fixed radius has the form
\be
\psi \sim t^{- (l + {n\over 2} - {1 \over 2})},
\label{osc1}
\ee
while the intermediate behaviour of SI fields at the outer horizon $H_{+}$ is dominated by an
oscillatory power law tails of the form as follows:
\be
\psi \sim v^{- (l + {n\over 2} - {1 \over 2})} \sin (mt).
\label{osc2}
\ee
In the next section we check our predictions numerically for various dimensions of the background spacetime.

\section{NUMERICAL RESULTS}
\begin{figure}
\begin{center}
\leavevmode
\epsfxsize=440pt
\epsfysize=540pt
\epsfbox{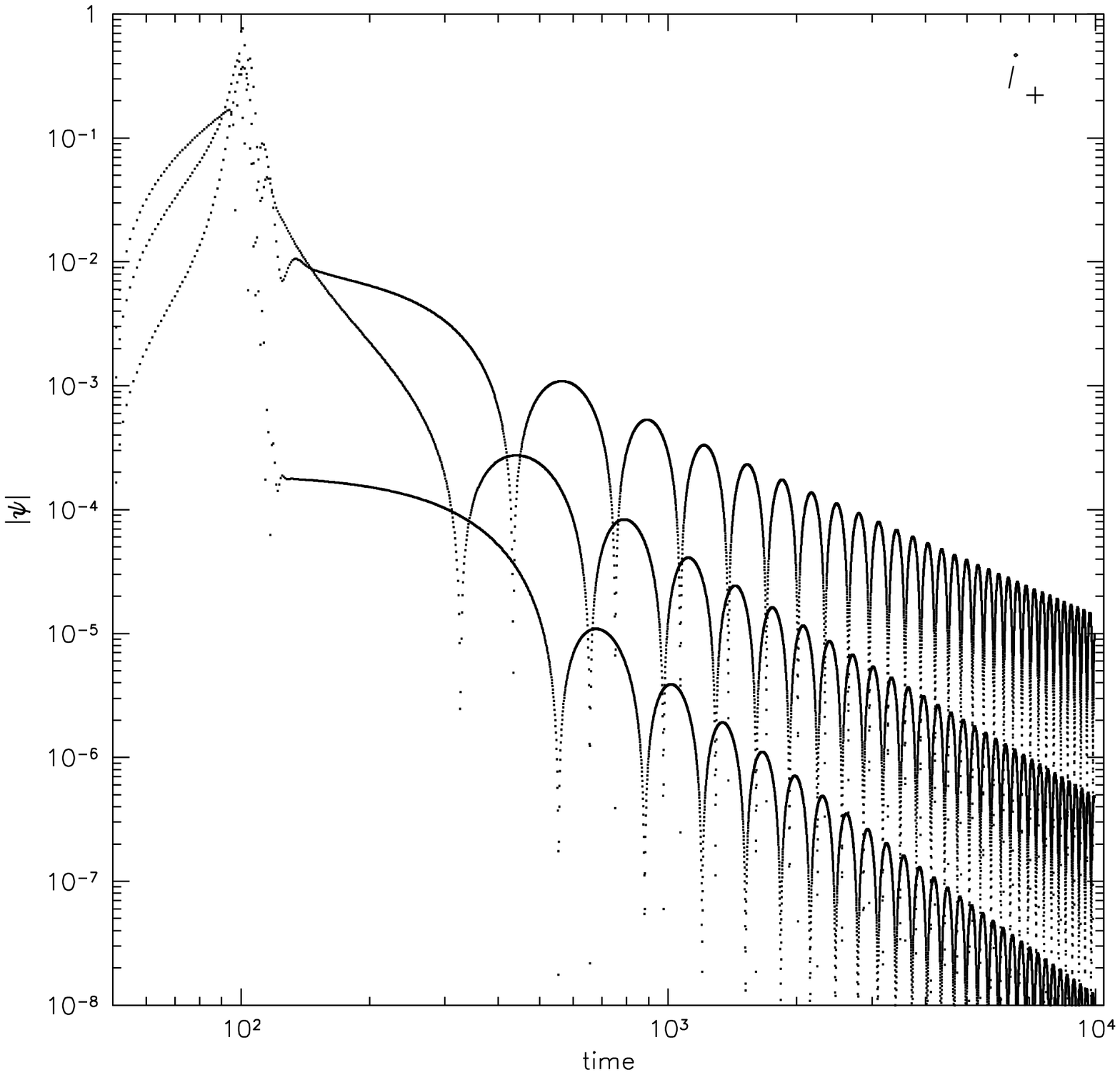}
\end{center}
\caption{Evolution of the field $|\psi|$ on the future timelike
  infinity $i_+$ (approximated by $\psi(y=50,t)$) as a function of t
  for different space dimensionality $n=4,5,6$ (curves from top to
  bottom, respectively). Calculation parameters are: $l=0$, $M=0.5$,
  $Q=0.45$, and $m=0.01$. The power-law exponents are $-1.51$,
  $-2.03$, and $-2.53$ for $n=4,5,6$, respectively. The period of the
  oscillations is $T = \pi/m \simeq 314.5 \pm 0.5$ to within $0.3\%$
  for all curves.}
\label{fig1a}
\end{figure}

We numerically analyzed Eq.~(\ref{mo}) using method described in
\cite{gun94}. We transformed Eq.~(\ref{mo}) into $(u,v)$ coordinates
\be
4 \psi_{,uv} + V \psi = 0,
\ee
and solved it on uniformly spaced grid using explicit difference
scheme.  As was pointed out previously the late time evolution of a
massive field is independent of the form of the initial data. In order to perform
our calculation we start with a Gaussian pulse of the form as follows:
\be
\psi(u=0,v) = A \exp \left ( - {(v-v_0)^2 \over \sigma^2} \right )
\ee
Because of the linearity of the relation (\ref{mo}) one has freedom in choosing
the value of the amplitude $A$. For our purpose we fix it as $A = 1$.
The rest of the initial field profile parameters we take as $v_{0} = 50$ and
$\sigma = 2$.\\
We shall set the mass of SI scalar field equal to $m=0.01$ and the mass
and charge of the black hole respectively equal to  $M=0.5$, $Q=0.45$. First, we shall
study the evolution of $\psi$ on the future timelike infinity $i_+$. In our calculations
we approximate this situation by the field at fixed radius $y = 50$. The numerical results
for $l = 0$ and different spacetime dimensions $n=4,5,6$ are shown in Fig.~\ref{fig1a}. Initially the evolution
is determined by the prompt contribution and quasinormal ringing. However, then with the
passage of time a definite oscillatory power-law fall off appears to be manifest.
We obtained the power-law exponents $-1.51$,
$-2.03$, and $-2.53$ for $n=4,5,6$, respectively. 
These values are to be compared with the analytically predicted ones equal respectively
to $-1.5$, $-2.0$, and $-2.5$. Thus, the agreement between numerical calculations
and analytically predicted values is excellent.
For the four-dimensional spacetime we
get the same result as obtained in Ref.\cite{hod98}. The period of the
oscillations is $T = \pi/m \simeq 314.5 \pm 0.5$ to within $0.3\%$
for all curves. Both values of power-law exponents and period of oscillation are
in perfect agreement with the analytical prediction.\\
Then, the evolution of the SI scalar field on the black hole future
horizon $H_+$ (approximated by $\psi(u=10^4,t)$), as a function of
$v$ for different space dimensionality $n=4,5,6$ was studied. Calculation parameters are: $l=0$, $M=0.5$,
$Q=0.45$, and $m=0.01$. The power-law exponents and the period of
the oscillations are the same as in Fig.~\ref{fig1a}.\\
Next, we take into consideration the dependence of SI scalar field on the multiple index.
We studied the evolution of the field $\psi$ on the future timelike
infinity $i_+$ as a function of $t$ for different multipoles $l=0,1$ and $2$ in
five-dimensional spacetime. The obtained power-law exponents are as follows: $-2.03$, and
$-3.03$, and $-4.04$ for $l=0,1,2$. 
According to relation (\ref{osc1}) these exponents are equal to $-2.0$, $-3.0$, and $-4.0$,
in perfect agreement with numerical calculations.
The period of the
oscillations is $T = \pi/m \simeq 314.5$ to within $1\%$ for all
curves, in agreement with the predicted value. The results are depicted in Fig.~\ref{fig2a}.
The same calculations were conducted for six-dimensional spacetime (Fig.~\ref{fig2b}), where
the power-law exponents are $-2.53$, and
$-3.53$, and $-4.53$ for $l=0,1,2$, respectively. 
Due to Eq.(\ref{osc1}) for six-dimensional spacetime they have the values:
$-2.5$, $-3.5$, and $-4.5$. Those values are also in agreement with numerical calculations.
The period of the
oscillations is $T = \pi/m \simeq 314.5$ to within $3\%$ (for the
worst case of $l=2$).\\
We also studied numerically the late-time behaviour of SI on black hole in five-dimensional spacetime, for
the field mass $m = 0.05$ (see Fig.~\ref{fig3a}). We investigated the behaviour on future timelike infinity and 
on black hole future horizon. The period of oscillation was $T = 63.0$ to within $0.8\%$. 
The field's
amplitude decays in agreement with the {\it no-hair theorem} due to the power-law fall off, contrary 
to the four-dimensional case where one has the decay rate slower than any power-law. In six-dimensional case
( Fig.~\ref{fig3b})
the behaviour of SI scalar fields is the same, with the period of oscillation equal to $T = 63.0$
to within $0.8\%$. The slope of the curve is equal to $-2.5$.\\
We also investigated the late-time behaviour on future timelike infinity in five-dimensional 
spacetime of static charged black hole for different masses of the scalar field Fig.~\ref{fig4a}.
The decay rates have the form of the power-law fall-off with the slope equal to $-2.0$. The same
studies were conducted in six-dimensional spacetime with the similar results Fig.~\ref{fig4b}, i.e.,
for various masses of SI scalar fields we obtained power-law fall off with the slope of the curve equal to $-2.5$.

\section{Conclusions}
We have elaborated analytically the intermediate behaviour of SI scalar fields in the background of
static charged $n$-dimensional black hole. It turned out that the intermediate asymptotic
behaviour did not depend on the spacetime parameters as $M$, $Q$ but only on the mass of SI scalar 
field. In other words, considering the SI scalar field perturbations one can neglect the
backscattering from the asymptotically far regions at intermediate times. We check our analytical calculations
by numerical ones and obtained excellent agreement with analytically predicted values.
Numerical studies of the late-time behaviour of SI scalar fields on the future timelike infinity
and on the future black hole horizon reveal the fact of the power-law decay (contrary to the four-dimensional case
where one has to do with the slower than any power-law decay). The same type of behaviour was obtained for various
masses of SI scalar fields in $n = 5, 6$ spacetime. This type of behaviour should be checked analytically, but due to the 
tremendous difficulties in solving differential equations is almost intractable. We hope to return to this problem
elsewhere.

\vspace{3cm}
\noindent
{\bf Acknowledgements:}\\
M.R. was supported in part by KBN grant 1 P03B 049 29.


\pagebreak

\begin{figure}
\begin{center}
\leavevmode
\epsfxsize=440pt
\epsfysize=540pt
\epsfbox{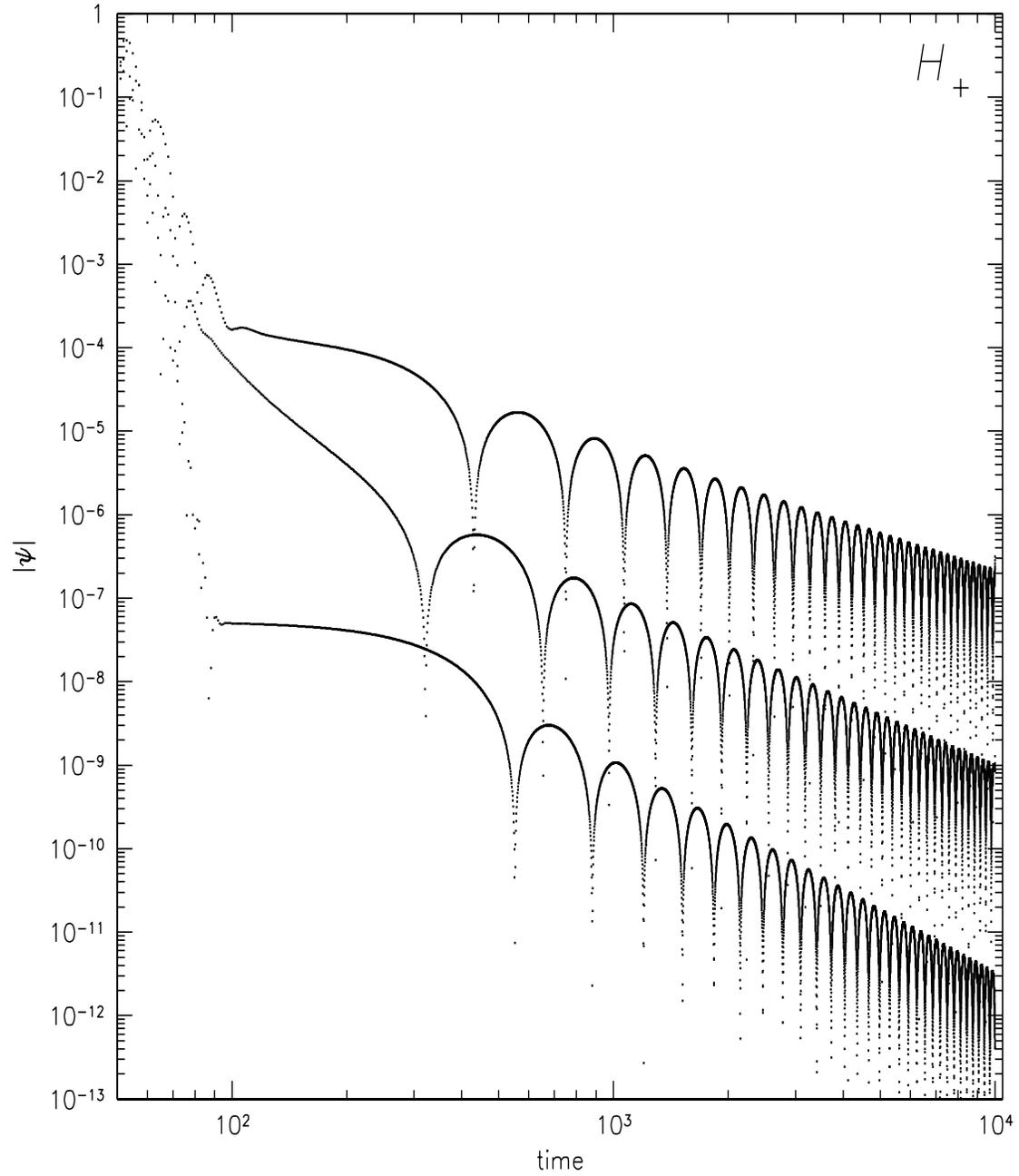}
\end{center}
\caption{Evolution of the field $|\psi|$ on the black hole future
  horizon $H_+$ (approximated by $\psi(u=10^4,t)$) as a function of
  $v$ for different spacetime dimensionality $n=4,5,6$ (curves from top to
  bottom, respectively). Calculation parameters are: $l=0$, $M=0.5$,
  $Q=0.45$, and $m=0.01$. The power-law exponents and the period of
  the oscillations are the same as in Fig.~\ref{fig1a}.}
\label{fig1b}
\end{figure}

\begin{figure}
\begin{center}
\leavevmode
\epsfxsize=440pt
\epsfysize=540pt
\epsfbox{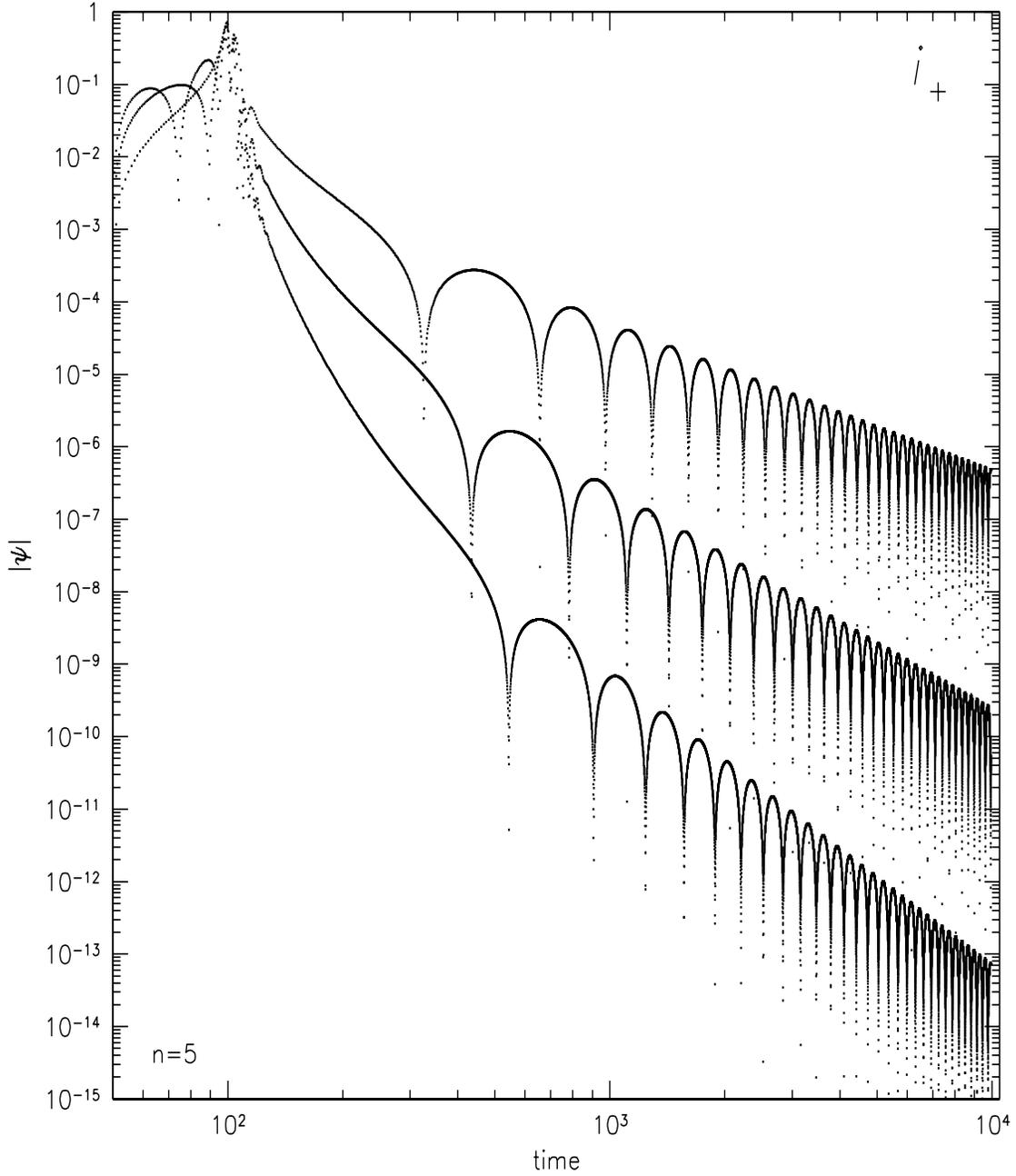}
\end{center}
\caption{Evolution of the field $|\psi|$ on the future timelike
  infinity $i_+$ (approximated by $\psi(y=50,t)$) as a function of t
  for different multipoles $l=0,1$ and $2$ (curves from top to bottom,
  respectively) and $n=5$. The power-law exponents are $-2.03$, and
  $-3.03$, and $-4.04$ for $l=0,1,2$, respectively. The period of the
  oscillations is $T = \pi/m \simeq 314.5$ to within $1\%$ for all
  curves. The rest of the parameters are the same as in
  Fig.~\ref{fig1a}.}
\label{fig2a}
\end{figure}

\begin{figure}
\begin{center}
\leavevmode
\epsfxsize=440pt
\epsfysize=540pt
\epsfbox{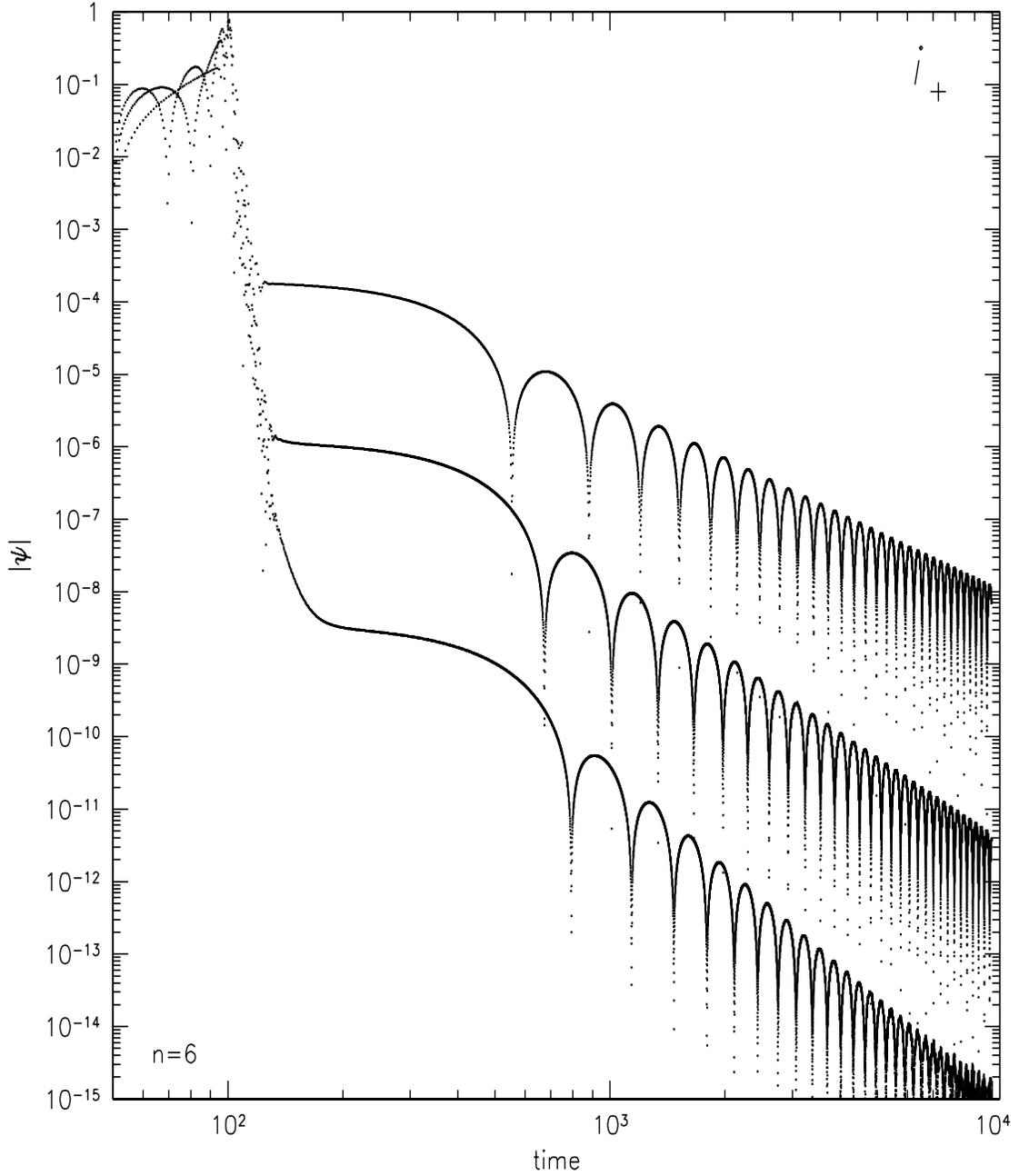}
\end{center}
\caption{Evolution of the field $|\psi|$ on the future timelike
  infinity $i_+$ (approximated by $\psi(y=50,t)$) as a function of t
  for different multipoles $l=0,1$ and $2$ (curves from top to bottom,
  respectively) and $n=6$. The power-law exponents are $-2.53$, and
  $-3.53$, and $-4.53$ for $l=0,1,2$, respectively. The period of the
  oscillations is $T = \pi/m \simeq 314.5$ to within $3\%$ (for the
  worst case of $l=2$). The initial data are those of Fig.~\ref{fig1a}.}
\label{fig2b}
\end{figure}
\begin{figure}
\begin{center}
\leavevmode
\epsfxsize=440pt
\epsfysize=540pt
\epsfbox{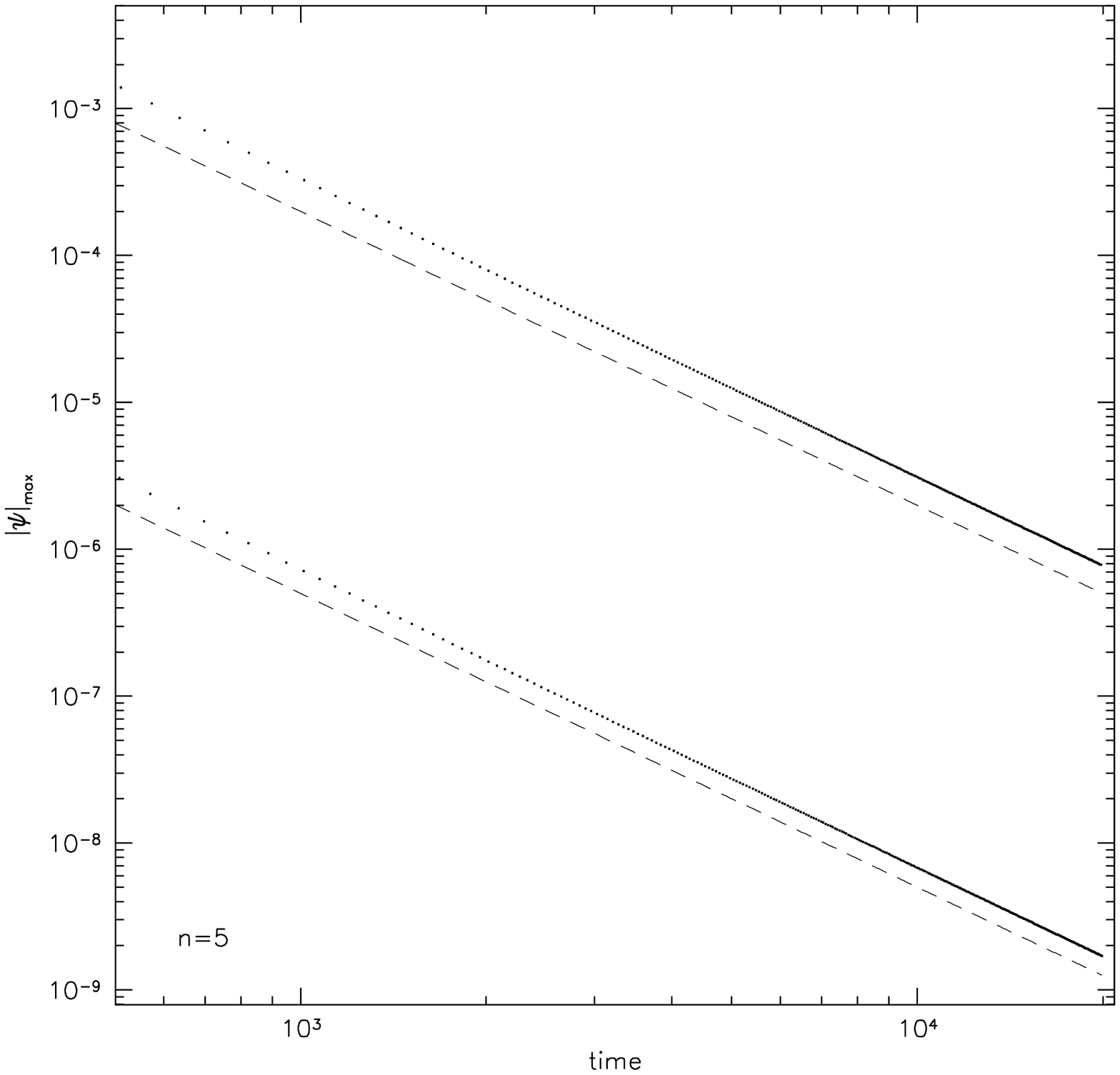}
\end{center}
\caption{Late-time behaviour of the field $|\psi|$ on black
 hole for $n=5$. Only maxima of the oscillations are shown. The mass of
 the field is $m=0.05$. The upper curve represents the field on future
 timelike infinity $i_+$ ($\psi(y=50)$) as a function of time
 $t$. Bottom curve is the field on the black hole future horizon $H_+$
 ($\psi(u=2 \times 10^4$) as a function of $v$. The period of the
 oscillations is $T = \pi/m \simeq 63.0$ to within $0.8\%$. The thin
 dashed lines have slopes equal to $-2.0$. The rest of the parameters are
 the same as in Fig.~\ref{fig1a}.}
\label{fig3a}
\end{figure}
 
\begin{figure}
\begin{center}
\leavevmode
\epsfxsize=440pt
\epsfysize=540pt
\epsfbox{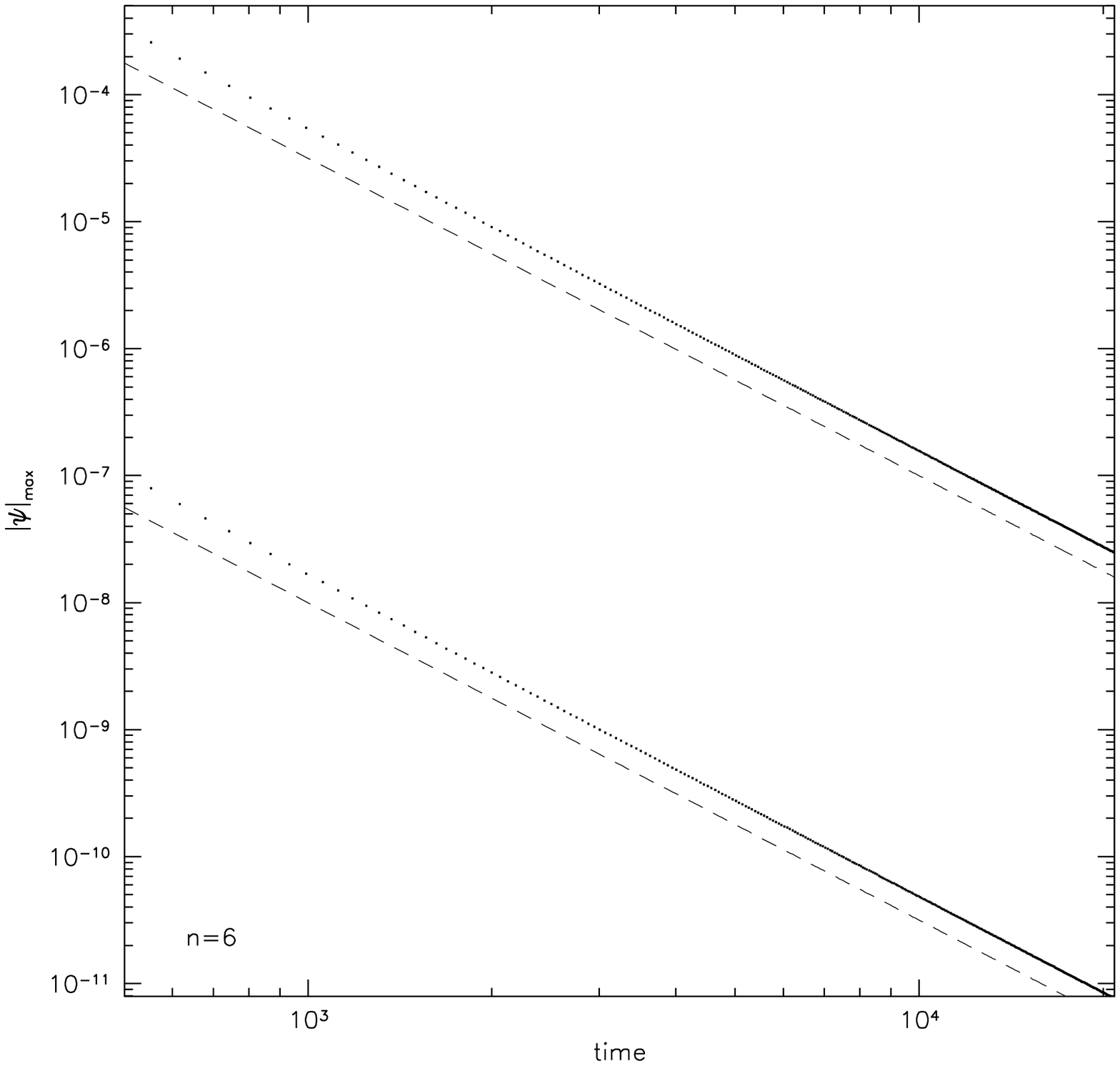}
\end{center}
\caption{Late time behaviour of the field $|\psi|$ on black
 hole for $n=6$. Only maxima of the oscillations are shown. The mass of
 the field is $m=0.05$. The upper curve represents the field on future
 timelike infinity $i_+$ ($\psi(y=50)$) as a function of time
 $t$. Bottom curve is the field on the black hole future horizon $H_+$
 ($\psi(u=2 \times 10^4$) as a function of $v$. The period of the
 oscillations is $T = \pi/m \simeq 63.0$ to within $0.8\%$. The thin
 dashed lines have slopes equal to $-2.5$. The rest of the parameters are
 the same as in Fig.~\ref{fig1a}.}
\label{fig3b}
\end{figure}

\begin{figure}
\begin{center}
\leavevmode
\epsfxsize=440pt
\epsfysize=540pt
\epsfbox{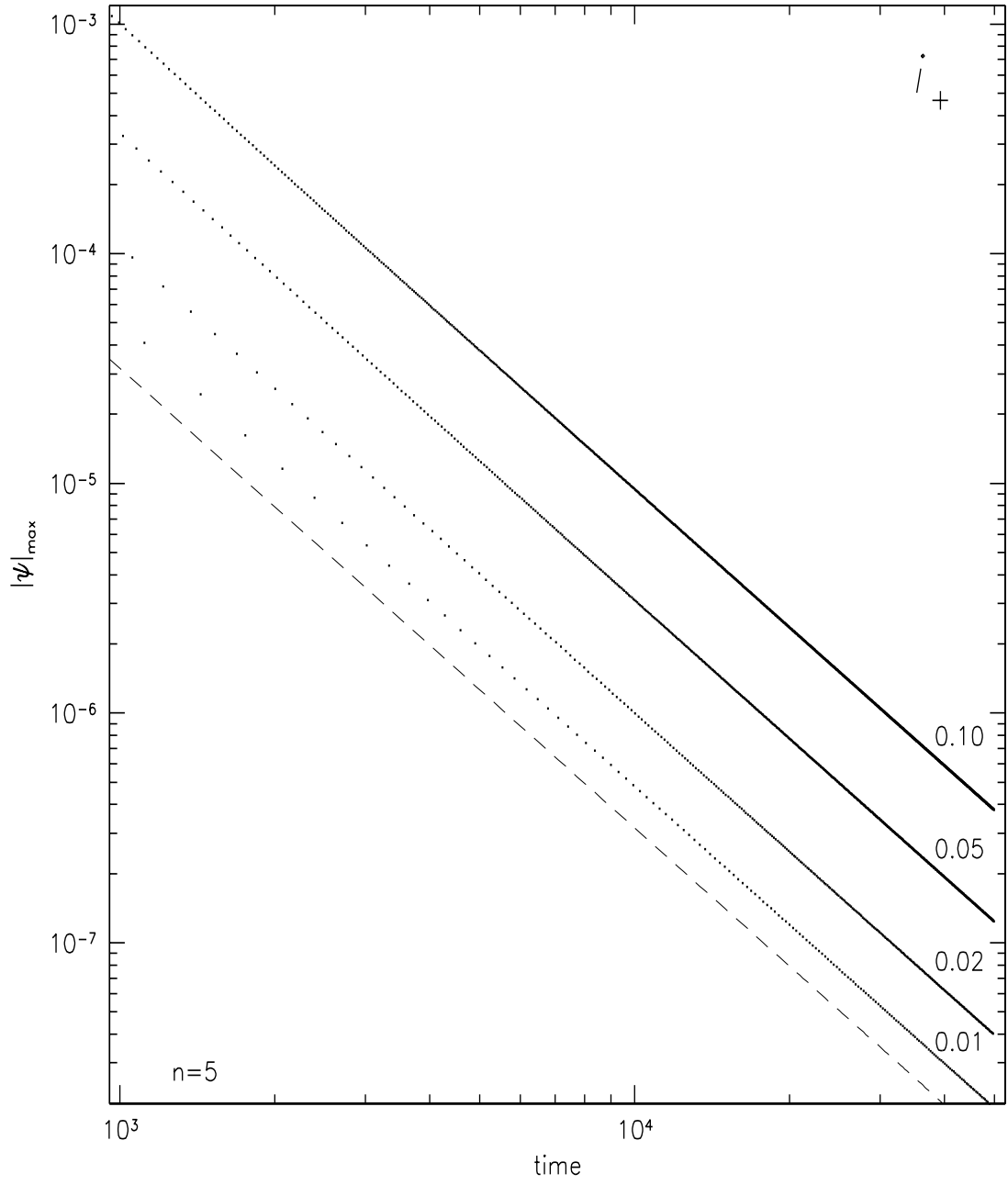}
\end{center}
\caption{Late-time behaviour of the field $|\psi|_{max}$ at future
 timelike infinity on black hole for $n=5$ and for different
 masses $m$. The dashed line has slope equal to $-2.0$. 
The initial data are those of Fig.~\ref{fig1a}.}
\label{fig4a}
\end{figure}

\begin{figure}
\begin{center}
\leavevmode
\epsfxsize=440pt
\epsfysize=540pt
\epsfbox{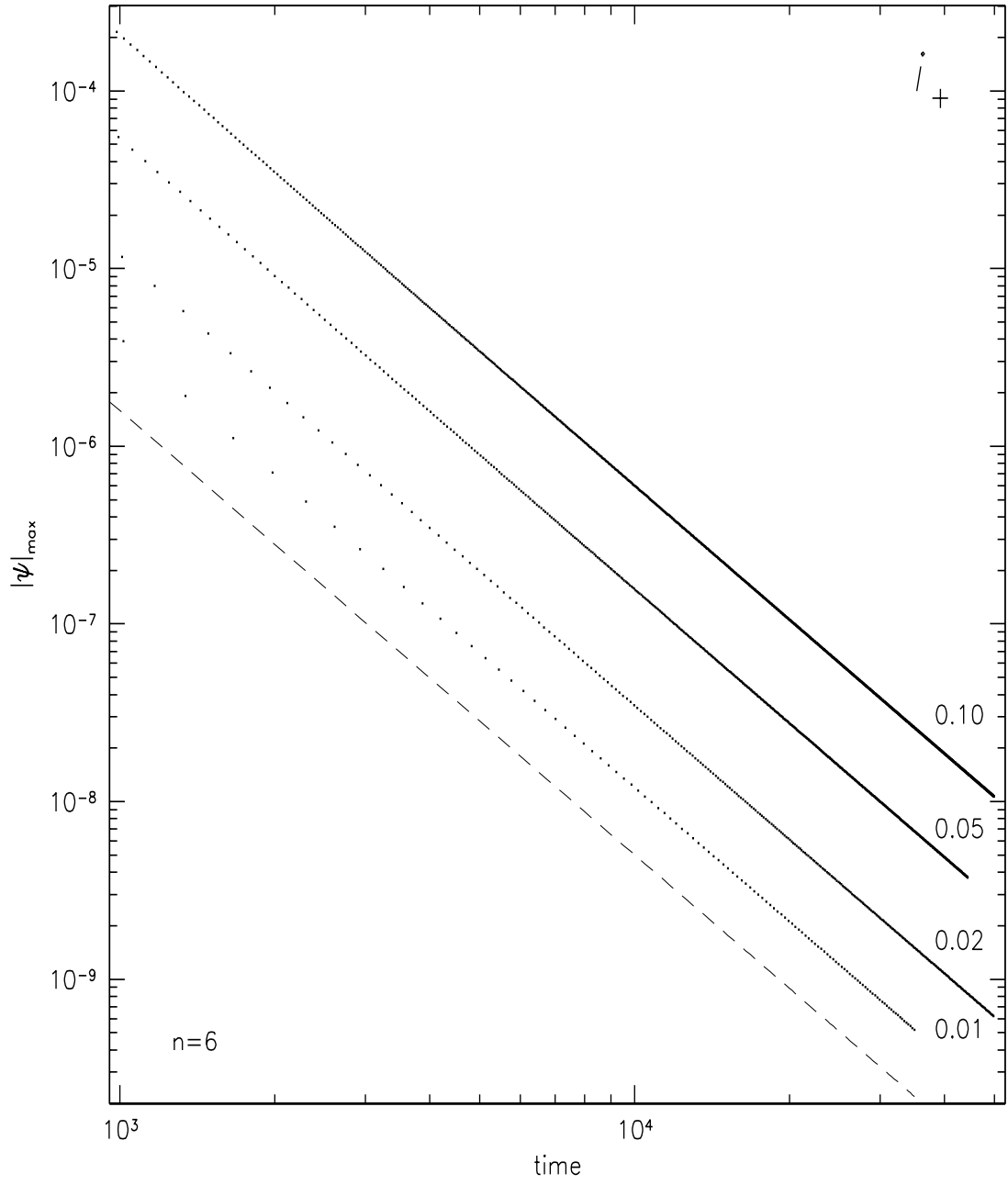} 
\end{center}
\caption{Late-time behaviour of the field $|\psi|_{max}$ at future
 timelike infinity on black hole for $n=6$ and for different
 masses $m$. The dashed line has slope equal to $-2.5$. The rest of the
 parameters are the same as in Fig.~\ref{fig1a}.}
\label{fig4b}
\end{figure}

\end{document}